\documentclass[aps,prl,twocolumn,showpacs,amsmath,amsfonts] {revtex4}
\usepackage{epsfig} \usepackage{bm} \usepackage{dcolumn} 
\renewcommand{\it}[1]{\textit{#1}}
\newcommand{\Ref}[1]{(\ref{eq:#1})}
\newcommand{\REF}[1]{Eq.~(\ref{eq:#1})}
\def\eg{{{e.g.}}} 
   \def\ie{{{i.e.}}}  
\def\vs{{{vs.~}}}
\def\nn {\nonumber} 
\def\br {\\ \nonumber} 
\newcommand{\B}[1]{{\bm{#1}}}
\newcommand{\C}[1]{{\mathcal{#1}}} 
\def\k{\B {k}} 
\def\d{\text{d}} 
 \newcommand{\ve}{\varepsilon}
\newcommand{\p}{\partial}      
\def\hf{\case{1}{2}}
\newcommand{\tk}{(t,\B k)} 
\newcommand{\pot}[1]{\frac{\partial {#1}}{\partial t}} 
\renewcommand{\sb}[1]{_{\text {#1}}}
 \begin{document}
\title{One-fluid description of  turbulently flowing suspensions}
\author{Victor S. L'vov and Anna Pomyalov}   
\affiliation{~Department of Chemical Physics, The Weizmann Institute
of Science, Rehovot 76100, Israel} 

\begin{abstract}
We suggested a \it{one-fluid} model of a turbulent dilute suspension
which accounts for the ``two-way'' fluid-particle interactions by
$k$-dependent effective density of suspension and additional damping
term in the Navier-Stokes equation. We presented analytical
description of the particle modification of turbulence including scale
invariant suppression of a small $k$ part of turbulent spectrum
(independent of the particle response time) and possible enhancemenent
of large $k$ region [up to the factor $(1+\phi)^{2/3}$]. Our results
are in agreement with qualitative picture of isotropic homogeneous
turbulence of dilute suspensions previously observed in laboratory and
numerical experiments.
\end{abstract}

\maketitle 

\textbf{Introduction}. The interaction of solid particles or liquid
droplets with turbulence in gases controls the performance of various
engineering devices like the combustion of pulverized coal and liquid
sprays and cyclone separations. This interaction plays an important
role in many areas of environmental science and physics of the
atmosphere. Dust storms, rain triggering, dusting and spraying for
agricultural or forestry purposes, preparation and processing of
aerosols are typical examples. For a review of turbulent flows with
particles and droplets see, \eg~\cite{98CST}.

In dilute suspensions with small volume fractions of particles, $C\sb
p$, the particle-particle interactions are negligible. Nevertheless,
for $\rho\sb p/\rho\sb f\gg 1$ (the ratio of the solid particle
material and the gas densities), the mass loading $\phi= C\sb p
\rho\sb p/ \rho\sb f$ may exceed unity and kinetic energies of the
particles and the carrier gas may be compatible; hence the ``two-way
coupling'': effect of fluid on particles and vice versa must be
accounted for. Current understanding of turbulence in dilute
suspensions is still at its infancy due to the highly nonlinear nature
of physically relevant interactions and a  wide spectrum of
governing parameters (the particle size $a$ \vs $L$ and $\eta$,
the outer and inner scales of turbulence, the particle response time
$\tau\sb p$ \vs $\gamma_{_L}$ and $\gamma_\eta$, turnover frequencies
of $L$- and $\eta$- scale eddies).

Analytical study of the problem is mainly based upon a \it{two-fluid}
model wherein both the carrying fluid and particle phases are treated
as interpenetrating continua~\cite{98CST,83EA,98BSS}. This model deals
with non-interacting solid spherical particles which are small enough
such that: (i) one can neglect the effect of preferential
concentration and assume homogeneity of the particle space
distribution; (ii) the Stokes viscous drag law for particle
acceleration, $d \B u\sb p/d t= [\B u\sb f-\B u\sb p]/\tau\sb p$, is
valid ($\B u\sb f$ is the fluid velocity).  Unfortunately, statistical
description of two-fluid turbulence by closure procedures requires a
set of additional questionable simplifications due to the lack of
understanding of the relevant physics of the particle-fluid
interactions. This made closures of the two-fluid model highly
qualitative at best~\cite{94Egl,98BSS,00BSS}.

We think that the basic physics of the problem may be described by a
simpler \it{one-fluid model} of turbulent dilute suspensions, which
requires standard closures of one-phase turbulence. The present Letter
suggests such a model and, as a first step, uses a properly modified
simple closure, based on the Kolmogorov-Richardson cascade picture of
turbulence.  The resulting non-linear differential equations for the
energy budget were solved analytically with required accuracy. This
provides an economical and internally consistent analytical
description of the turbulence modification by particles including the
dependencies of suppression or enhancement of the turbulent activity
on the three governing parameters: $(\tau\sb p \gamma_{_L})$, $\phi$
and scale of eddies.  These effects were previously observed in
numerous experimental and numerical
studies~\cite{98CST,83EA,94Egl,98BSS,00BSS,00HKR}, but they still await
analytical description.  Our analytical findings are in a qualitative
agreements with the observations. We believe that the one-fluid model,
together with more elaborated closures of one-phase
turbulence~\cite{02LOP}, offers an insightful and constructive look at
basic physics of even more complex particle-laden turbulent flows.

\noindent
\textbf{1. One-fluid model of turbulent suspensions} reads:
\begin{eqnarray} \label{eq:NS}
 && \rho\sb{eff}(k) \Big [\pot{}+ \gamma\sb p (k)\Big] 
\B u\tk + \mu k^2
 \B u\tk  \br &&+ (2\pi)^{-3}\int \d^3 k_1\, \d^3 k_2 
 \Gamma^{\alpha\beta\gamma}_{\B k\,\B k_1\B k_2} u_\beta^*(t,\k_1)
 u_\gamma^* (t,\k_2) =0\ .
\end{eqnarray} 
Here $\B u\tk$ is the incompressible velocity field of the carrier
fluid in the $\k$ representation and $\mu$ is the dynamical
viscosity. This equation differs from the Navier-Stokes (NS) equation
in the three aspects:

$\B a$. The carrier fluid density $\rho\sb f$ is replaced by
$\rho\sb{eff}(k)$, a $k$-dependent effective density of the suspension
for turbulent fluctuations of scale $1/k$ [referred to as $k$-eddies];

$\B b$. The fluid-particle friction is described by a damping
frequency $\gamma\sb p (k)$.

$\B c$. The interaction amplitude of the NS equation,\\
$ \gamma^{\alpha\beta\gamma}_{\B k\,\B k_1\B k_2}=\rho\sb f
\big[P^{\alpha\beta}(\B k)\,k^\gamma +P^{\alpha\gamma}(\B k)\,
k^\beta\big] \delta(\B k+\B k_1+\B k_2)/2$,
$P^{\alpha\beta}(\k)=\delta_{\alpha\beta}-k^\alpha k^\beta/k^2$ is
replaced by
\begin{equation} \label{eq:int2}
\Gamma^{\alpha\beta\gamma}_{\B k\,\B k_1\B k_2}= \rho\sb
{eff}\big[2\,k_1k_2k_3/(k_1^2+k_2^2+k_3^2)\big]
\gamma^{\alpha\beta\gamma}_{\B k\,\B k_1\B
k_2}\big /\rho\sb f \ .
\end{equation}
An underlying physics of these aspects is quite simple:

$\B a$. For small $k$ the turnover frequency $\gamma(k)$ of
$k$-eddies is small in the sense $\gamma(k)\tau\sb p\ll 1$. Therefore
in this region of $k$ the particle velocity is very close to that of
the carrier fluid and we can describe the suspension as \it{one fluid}
with effective density $\rho\sb {eff}(k)$ which is close to the
density of suspension, $\rho\sb s=\rho\sb f+ C\sb p \, \rho\sb p$.  In
contrary, for large $k$, when $\gamma(k)\tau\sb p\gg 1$, the
particles cannot follow these very fast motions and may be considered
at rest. Thus the particles do not contribute to the effective
density and $\rho\sb {eff}(k)\to \rho\sb f$. In general case $\rho\sb
{eff}(k)$ may be given by
\begin{eqnarray}
  \label{eq:eff-den}
   \rho\sb {eff}(k)=\rho\sb f \big[1+ \phi f\sb{com}(k)]\,,\quad
   \phi=C\sb p\,\rho\sb p/ \rho\sb f\ .
\end{eqnarray}
Here a statistical ensemble of all particles, partially involved in
the motion of $k$-eddies, is replaced by two sub-ensembles of ``fully
comoving'' fraction $f\sb{com}(k)$ of particles and ``fully resting''
fraction $ f\sb{rest}(k)=1-f\sb{com}(k)$ of particles, which does not
contribute to $\rho\sb{eff}(k)$.

$\B b$. The particles at rest cause fluid-particle friction. According
to Newton's third law, the damping frequency of suspension $\gamma\sb
p(k)$ may be related to the particle response time, $\tau\sb p$, via
the ratio of total mass of particles $M\sb p$ to the total effective
mass of the suspension $M\sb{eff}(k)$:
\begin{eqnarray}
\label{eq:gamma-p}
\gamma\sb p(k)=\frac{M\sb p}{\tau\sb p \, M\sb{eff}(k)}
= \frac{ C\sb p\, \rho\sb p
f\sb{rest}(k)}{\tau\sb p\, \rho\sb {eff}(k)}
= \frac{\phi\,  \rho\sb f \, f\sb{rest}(k)\,
}{ \tau\sb p\, \rho\sb {eff}(k)}\ .
\end{eqnarray}
As we mentioned, the portions $f\sb{com}(k)$ and $f\sb{rest}(k)$
depend on $\tau\sb p\gamma(k)$. Clearly, for
$\tau\sb p\gamma(k)\ll 1$, $f\sb {rest}(k)$ has the same smallness:
$f\sb {rest}(k)\sim \tau\sb p\gamma(k)$. In the opposite case when
$1/\tau\sb p\gamma(k)$ is small, $f\sb{com }(k)$ has corresponding
smallness: $f\sb{com}(k)\sim 1/\tau\sb p\gamma(k)$. For $\tau\sb
p\gamma(k)=1$ one expects $f\sb {rest}(k)\simeq f\sb {com}(k)\simeq
\hf$. As a simple model of such a function we adopt
\begin{eqnarray} \label{eq:rest}
  f\sb {rest}(k)&=& 1- f\sb {com}(k) =\tau\sb p\gamma(k)/ [ 1+
  \tau\sb p\gamma(k) ]\,,
\end{eqnarray}
which also follows from more elaborated analysis~\cite{02LOP}.  With
\REF{rest} we rewrite Eqs.~\Ref{eff-den} and \Ref{gamma-p} as follows:
\begin{eqnarray}\label{eq:par1}
\rho\sb{eff}(k)&=&\rho \sb f \big\{1+\phi\big / [1+ \tau\sb p
\gamma(k)]\big\} \,,\\ \label{eq:par2} 
\gamma\sb p(k)&=&\phi\, \gamma(k)\big /
[1+ \phi+ \tau\sb p \gamma(k)]\ .
\end{eqnarray}

$\B c$.  It may be shown that interaction amplitude
$\Gamma^{\alpha\beta\gamma} _{\B k\,\B k_1\B k_2} $ is \it{Galilean
invariant} and \it{conserves the total kinetic energy} of suspension
(\ie ~of the carrier fluid and fully comoving part of the particles)
if one neglects the damping terms $\mu k^2$ and $\gamma\sb p(k)$ in
\REF{NS}. Notice, that the detailed form of
$\Gamma^{\alpha\beta\gamma} _{\B k\,\B k_1\B k_2} $ is not important
for this discussion, only the conservation of the energy is essential.

The basic equations of our model, \Ref{NS}, \Ref{int2}, \Ref{par1} and
\Ref{par2} are \it{self-consistent equations for turbulent velocity of
the carrier fluid} in which the coefficients depend on the
eddy-turnover frequency $\gamma(k)$ which, in its turn, depends on a
stochastic solution of the same equations.

\noindent
\textbf{2. Budget of the kinetic energy in suspensions}.  The
definition of one-dimensional density of kinetic energy in a
\it{single phase flow} $E(t,k)$ reads
\begin{eqnarray}
  \label{eq:energy1}
E(t,k)&=& \rho\sb f k^2 F_2(t,k)/2\pi\ .
\end{eqnarray}
Here $ F_2(t,k)$ is the simultaneous pair velocity correlation for
isotropic turbulence. Corresponding definition of one-dimensional
density of kinetic energy \it{for suspensions}, $\C E(k)$, has to
account for the $k$-dependent density:
\begin{eqnarray}
  \label{eq:energy2}
\C E(t,k)&=& \rho\sb {eff}(k)k^2 F_2(t,k)/2\pi\ .
\end{eqnarray}
Multiplying \REF{NS} by $\B u(t,\k')$ and averaging, one gets the
 equation for the energy budget in the inertial interval:
\begin{eqnarray}\label{eq:budget}
\C E(t,k)/2 \,\p t+\gamma\sb p(k) \C E(t,k) + \d
  \ve(k)/\d k=0\ .
\end{eqnarray}
Here $\ve(k)$ is \it{one-dimensional energy flux} in the $k$-space and
we neglected the energy input in the outer scale $L$ and the viscous
damping, $\mu k^2$, which becomes essential near the viscous
microscale $\eta$.

In the Richardson-Kolmogorov picture of turbulence the only relevant
parameters in the inertial interval are $k$, $\rho\sb f$ and
$\ve(k)$. Similarly, in our model~\Ref{NS} these parameters are $k$,
$\rho\sb {eff}(k)$ and $\ve(k)$. Other functions in the problem may be
related to them by the dimensional reasoning:
\begin{eqnarray}
  \label{eq:K41e}
\C E(k) &=&C_1 \, \big[ \varepsilon
(k)^2\rho\sb{eff}(k)\big]^{1/3}k^{-5/3}\,, \\ \label{eq:K41g}
\gamma(k)&=&C_2\, \big[\ve(k)\big/\rho\sb{eff}(k)\big]^{1/3}k^{2/3}\,,
\end{eqnarray}
where $C_1\sim C_2\sim 1$ are dimensionless constants for particle
free case. Hereafter we omit the explicit reference to the time
dependence.  Substituting Eqs.~\Ref{par1}, \Ref{par2}, \Ref{K41e}
and \Ref{K41g} into~\REF{budget} yields in the stationary case
\begin{eqnarray}\label{eq:bud}
\frac{\d \ve(k)}{\d k}+ \frac{\ve(k)}{k}\, \frac{C \,\phi}{1+\phi+
\gamma(k)\tau\sb p}=0\,, \quad C=C_1\,C_2\ .
\end{eqnarray}
To find the iterative solution of \REF{bud} we denote as
$\ve_n(k)$, $\gamma_n(k)$ and $\rho_{\text{eff},n}(k)$ corresponding
functions at $n$th iteration step; take for ``zeroth step'' their
values at $k\sb{outer}=1/L$: $\ve_0(k)=\ve(1/L)\equiv \ve_{_L}$,
$\gamma_0(k) =\gamma(1/L) \equiv \gamma_{_L}$, $\rho_{\text{eff},0}(k)
=\rho\sb{eff}(1/L)\equiv \rho_{_L}$; define dimensionless functions of
$\kappa\equiv kL$: $\ve_n(k)=\ve_{_L}f_n(\kappa)$,
$\gamma_n(k)=\gamma_{_L} g_n(\kappa)$, $\rho_{\text{eff},n}(k)
=\rho\sb f\,r_n(\kappa)$ and iterate the equations
\begin{eqnarray}\nn
 f_n(\kappa)&=&\exp\left\{ \int _1^\kappa  \frac{-\Delta \, \d x}
{x \left[1+\delta g_{n-1}(x) \right]}\right\}\,,
\quad   \delta=\frac{\tau\sb p\gamma_{_L}}{1+\phi}\,, \br 
g_n(\kappa)&=&\Big[\frac{\kappa^2 
f_n(\kappa)r_{n-1}(1)}{r_{n-1}(\kappa)}\Big]^{1/3}\,,
\hskip 1.2cm \,
\Delta =\frac{C\,\phi}{1+\phi}\,, \\ 
r_n(\kappa)&=&1+\phi\big/ [1+\tau\sb p
\gamma_{_L}\, g_n(\kappa)]\,, \label{eq:iter}
\end{eqnarray}
which coincide with~\Ref{bud}, \Ref{K41g}, \Ref{par2} after
ignoring subscript ``$_n$''. First two iterations may be done
explicitly:
\begin{eqnarray}\nn
f_1(\kappa)&=&\frac{\ve_1(k)}{\ve_{_L}}=\kappa^{-\Delta}\,,\quad
g_1(\kappa)=\frac{\gamma_1(k)}{\gamma_{_L}}=\kappa^{(2-\Delta)/3}\,,
\\ \label{eq:sol1} 
r_1(\kappa)&=&\frac{\rho_{\text{eff},1}(k)}{\rho\sb f}
=1+\frac{\phi}{1+\tau\sb
p\gamma_{_L}\kappa^{(2-\Delta)/3}}\,;
\\ 
 f_2(\kappa)&=&\frac{\ve_2(k)}{\ve_{_L}}=\Big[\frac{\delta +
 \kappa^{(\Delta-2)/3}}{\delta+1} \Big] ^{3 \Delta /(2-\Delta)}\,,
 \label{eq:sol2}\br 
g_2(\kappa) &=&\gamma_2(k)/\gamma_{_L}
 =\kappa^{2/3}f_2^{1/3}(\kappa)r_1^{-1/3}(\kappa)\,,\br
 r_2(\kappa)&=&\rho_{\text{eff},2}  (k)/\rho\sb f
=1+\phi\big/ [1+\tau\sb p \gamma_{_L}\, g_2(\kappa)]\ .
\end{eqnarray} 
The third iteration requires simple numerical integration:
\begin{eqnarray}\label{eq:sol3}
 \ve_3(k)=\ve_{_L}\exp\Big\{ -\Delta \, \int _1^{kL}\d x/ x
 \left[1+\delta g_2(x) \right]\Big\}\ .
\end{eqnarray} 
\begin{table}
\caption{\label{tab:1} Parameters  $\phi$ and $\tau\sb p$ for
  numerical solution of Eqs.~\Ref{K41e} -- \Ref{bud} with
  $C_1=C_2=\hf$, and computed values of $\delta$, crossover scale,
  $k_*$, and normalized rates of dissipation: by the iteration
  procedure, $f_{2\infty}$, $f_{3\infty}$ and numerically,
  $f_\infty$.}
\begin{ruledtabular}
\begin{tabular}{||c|rrrrr|rrr||}
$\phi$& & & $1.0$ & & &$0.75$&$0.5$&$0.25$\\ \hline $\tau\sb
p$&$1.0$&$0.5$&$0.2$&$0.1$&$0.02$& &$0.5$& \\ \hline\hline    
$\delta$&$0.21$&$0.10$&$0.04$&$0.02$&$0.004$&$0.12$&$0.15$&$0.19$\\
$\tau\sb p \gamma_{_L}$&
$0.42$&$0.20$&$0.08$&$0.04$&$0.008$&$0.21$&$0.22$&$0.24$\\
$k_*$&$3.4$&$9.5$&$39$&$117$&$1530$&$9.2$&$8.8$&$8.4$\\ \hline\hline
$f_{2,\infty}$&$0.704$&$0.621$&$0.522$&$0.455$&$0.331$
&$0.686$&$0.766$&$0.868 $\\
$f_{3,\infty}$&$0.721$&$0.639$&$0.537$&$0.469$&$0.341$
&$0.699$&$0.774$&$0.871 $\\
$f_\infty$&$0.723$&$0.642$&$0.541$&$0.474$&$0.352$&$0.701$&$0.775$
&$0.871$\\
\end{tabular}
\end{ruledtabular}
\end{table}
\begin{figure}
\epsfxsize=7.4cm 
\epsfbox{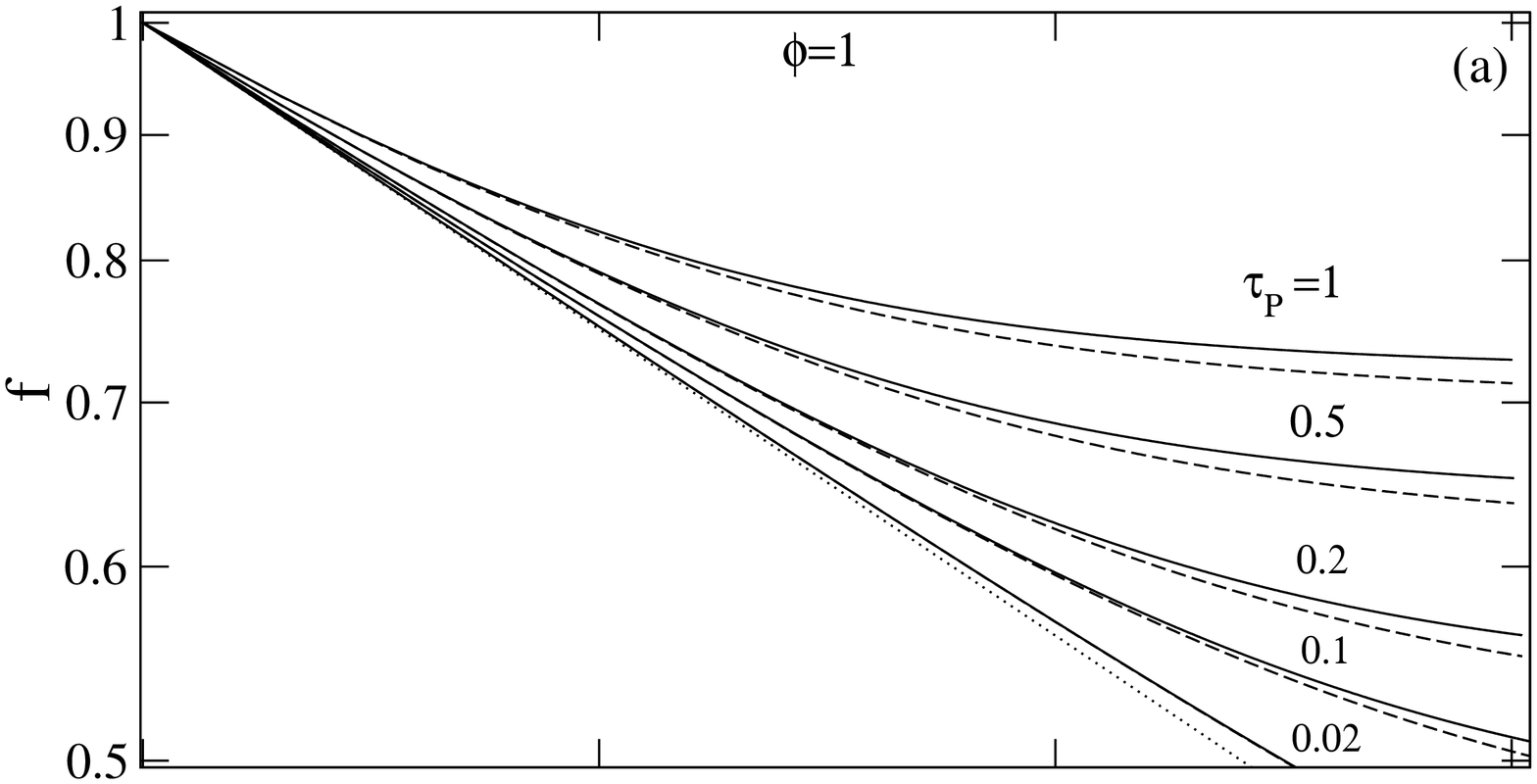}
\epsfxsize=7.6cm \epsfbox{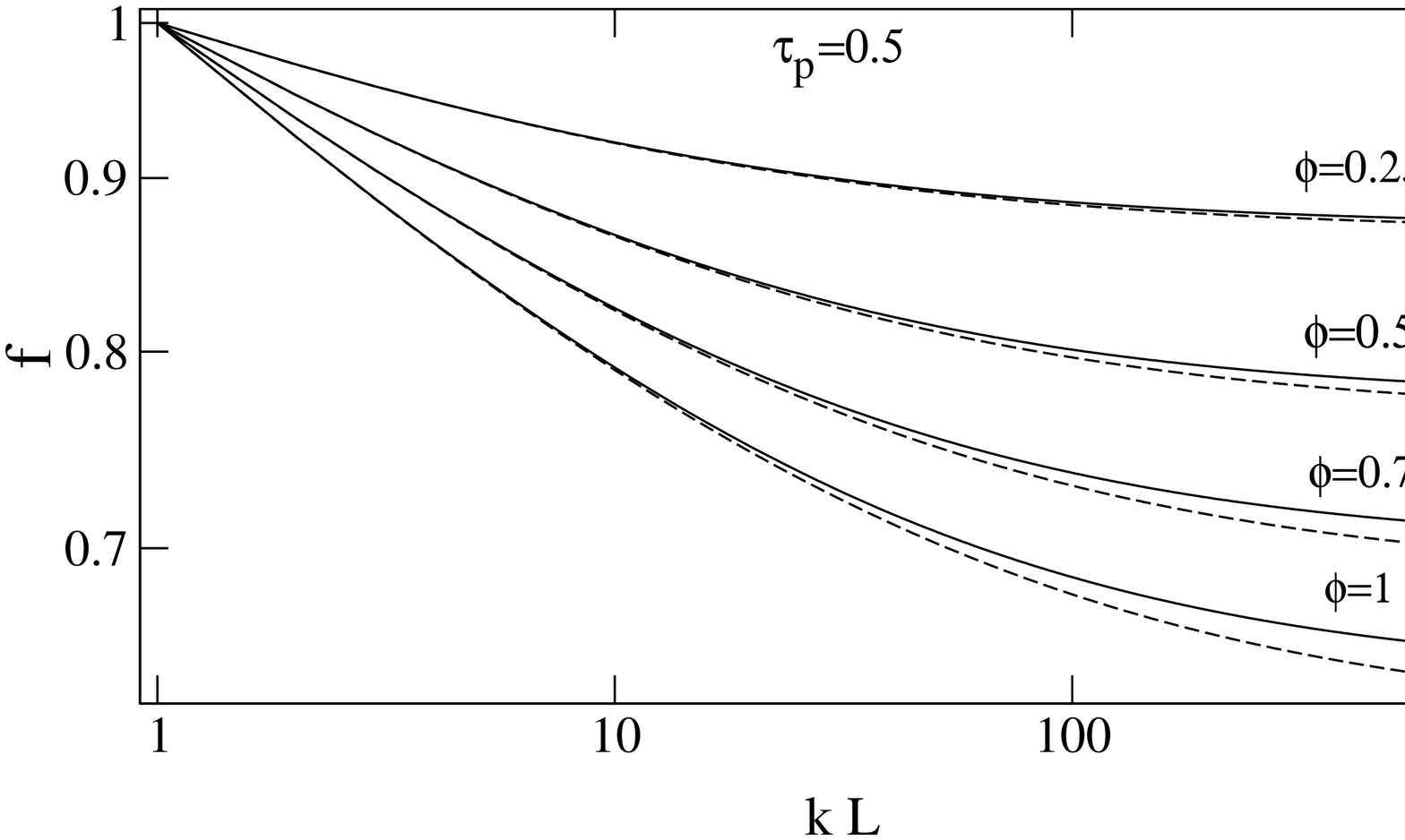}
\caption{\label{f:1}Log-Log plots of solutions of Eqs.~\Ref{K41e} --
  \Ref{bud} for parameters in Tab.~\ref{tab:1}. Solid lines: ``exact''
  numerical solutions $f(kL)=\ve(k)/\ve_{_L}$, dashed lines:
  $f_2(kL)=\ve_2(k)/\ve_{_L}$, \REF{sol2}. }
\end{figure}

Figure~\ref{f:1} displays ``exact'' numerical solutions of
$\ve(k)/\ve_{_L}$ (for sets of $\phi$ and $\tau\sb p$ listed in
Tab.~\ref{tab:1}) in comparison with corresponding plots of
Eq.~\Ref{sol2} for $f_2(\kappa)$.  In the first decade of the inertial
interval ($kL\le 10$) discrepancies between the numerics and the
second iteration are quite small. They gradually increase toward large
$k$, remaining smaller than $(1-2)\%$.  Values of $f_\infty\equiv
f(\infty)$, $f_{2,\infty}=\left[\delta /
(1+\delta)\right]^{3\Delta/(2-\Delta)}$ and $f_{3,\infty}$ following
from \REF{sol3} are given in Tab.~\ref{tab:1}. It is clear that almost
always one may use simple analytical solution of the second
iterations, \Ref{sol2}.  The results \Ref{sol3} of the third iteration
may be used to control accuracy.

\noindent
\textbf{3.~Turbulence modification by particles}. In the particle-free
turbulence the rate of energy input $\ve_{_L}$ is equal to the energy
flux in the inertial interval $\ve(k)$ and to the rate of energy
dissipation $\ve_\infty$. In turbulent suspensions due to
fluid-particle friction $\ve(k)$ is no longer constant and decreases
toward large $k$. Therefore $\ve_{_L}>\ve(k)>\ve_\infty$.  \REF{sol1}
and Fig.~\ref{f:1}a show that in the region $k\ll k_*$ [$\tau\sb p
g(k_*)=1$] the flux decays toward small scale: $\ve(k)\approx
\ve_{_L}(kL)^{-\Delta}$ with $\Delta$, \REF{iter}, independent of
$\tau\sb p$, while for $k\gg k_*$ the flux approaches plato, similar
to the particle-free case. At first glance these two facts are
unexpected: there is an essential suppression of the large scale
eddies in spite of the fact that particles are almost swept by them
and therefore the fluid-particle damping $\gamma\sb p(k)$ is small. On
the contrary, the small scale eddies are almost not effected by
particles which are not involved in their motion and thus $\gamma\sb
p(k)$ reaches the maximum value $\phi/\tau\sb p$. To explain consider
the dimensionless rate of the flux decrement, $[- d \ln \ve(k)/ d \ln
k]$, which is $\propto \gamma\sb p(k)$. The only available
frequency to make $\gamma \sb p(k)$ dimensionless is $\gamma(k)$.
Therefore $- d \ln \ve(k)/ d \ln k \sim \Gamma(k)\equiv \gamma\sb
p(k)/\gamma (k)$, in agreement with \REF{bud}. As one sees from
\REF{par2}
\begin{eqnarray}
  \label{eq:Gamma}
\Gamma(k)=\phi/[ 1+\phi+\tau\sb p \gamma(k)]
\end{eqnarray}
and for $\tau\sb p \gamma(k)\ll 1$ the ratio $\Gamma(k)$ becomes
$\tau\sb p$ independent constant. This explains both facts: $\B a$ -
why the suppression of small $k$ eddies is $\tau\sb p$ independent and
$\B b$~- why this suppression is scale invariant. Note that the weak
sensitivity of turbulent spectra to $\tau\sb p$ (fact $\B a$) was
previously observed in numerous simulations of turbulence in dilute
suspensions but, to the best of our knowledge, was not well
understood, see, \eg~\cite{98BSS}. Equation~\Ref{Gamma} shows that
$\Gamma(k)\to 0$ for $\tau\sb p \gamma(k)\gg 1$.  Therefore in this
region of $k$ particles cannot modify the turbulence and indeed
$\ve(k)$ must approach a constant value, $\ve_\infty$.

\begin{figure}
\epsfxsize=7cm 
\epsfbox{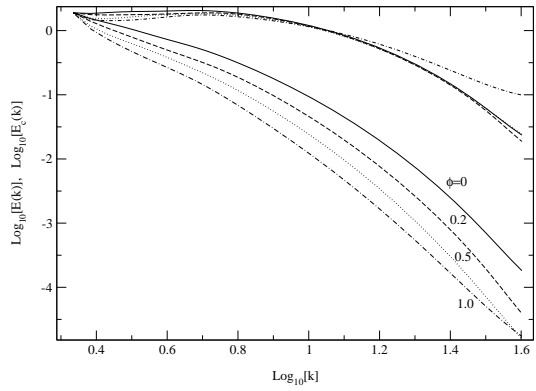}
\caption{\label{f:exp} Log-Log plots of turbulent kinetic 
energy spectrum $E(k)$ taken from \cite{98BSS} for 
$\phi=0, 0.2, 0.5$ and 1 (lower lines) and
``compensated'' spectra $E\sb c(k)$ for the same $\phi$ (upper
lines).}
\end{figure}

For brief comparison of the prediction $\B b$ with direct numerical
simulation by Boivin, Simonin and Squires~\cite{98BSS}) we reploted in
Fig.~\ref{f:exp} their Fig.~5b for kinetic energy spectra $E(k)$ of
suspensions in Log-Log coordinates (lower lines). Our first
iteration~\Ref{sol1} predicts $E(k)\propto k^{-5/3-2\Delta/3}$ with
$\Delta=C\phi/(1+\phi)$. Therefore we defined ``compensated'' spectra
as $E\sb c(k,\phi)=E(k)(kL)^{5/3+2\Delta/3}$ and plotted them in
Fig.~\ref{f:exp} (with $C=3.8$). The region $0.4<$Log $k<1$ where for
$\phi=0$, $E\sb c(k)=E(k)(kL)^{5/3}$, upper solid line, is
approximately constant may be considered as inertial interval.  As we
expected, in this interval all lines for different $\phi$ are about to
collapse. Some scattering of the lines is related with $k$-dependence
of $\rho\sb{eff}(k)$ and finite value of $\tau\sb p$ which is
neglected in the first iteration. More detailed comparison of our
second and next order iterations will be done elsewhere~\cite{02LOP}.

\begin{figure}
\epsfxsize=7.5cm 
\epsfbox{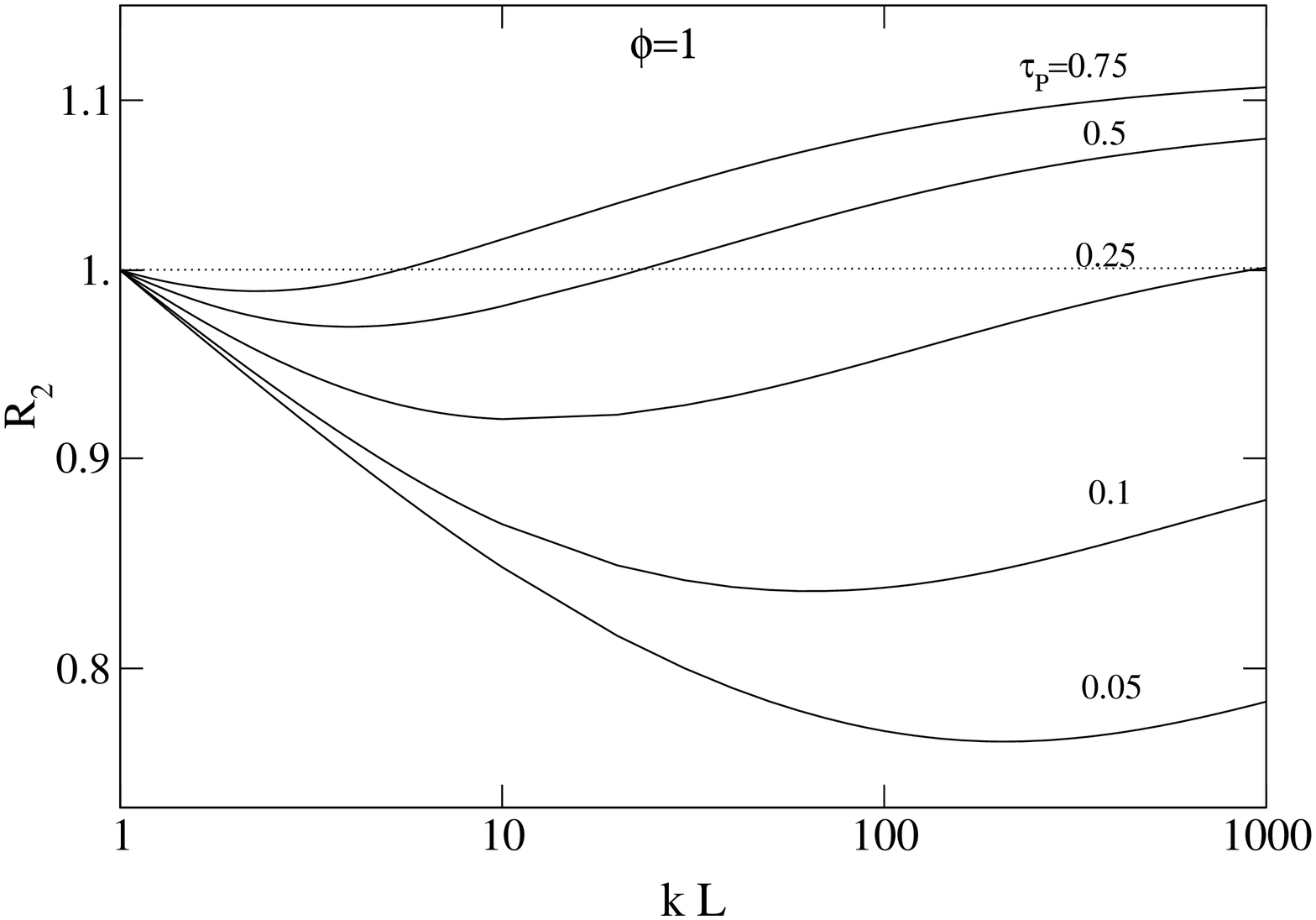}
\epsfxsize=7.2cm 
\epsfbox{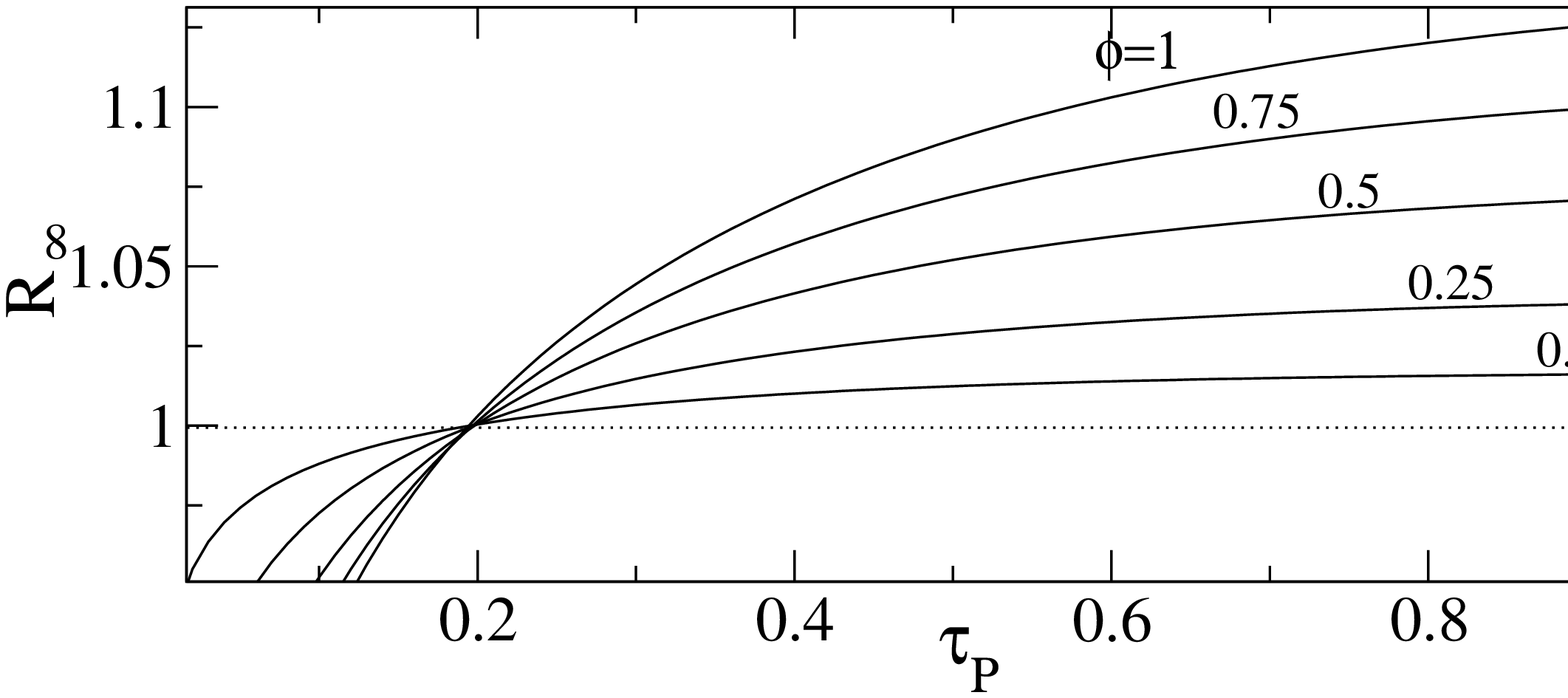}
\caption{\label{f:3} Upper panel: Log-log plots of functions $R_2(k)
=[f_2(\kappa)r_2(1)/r_2(\kappa)]^{2/3}$, Eqs.~\Ref{sol2},\,
\Ref{ratio}, for $\phi=1$ and values of $\tau\sb p$ denoting
corresponding lines. Lower panel: Plots of $R_{2,\infty}$, \REF{R2inf},
 \vs $\tau\sb p$. Values of $\phi$ label corresponding lines. }
\end{figure}

Consider now a possible enhancement of the density
of kinetic energy of the carrier fluid $E(k)$.  According to Eqs.
\Ref{energy1}, \Ref{energy2} and \Ref{K41e}
$
E(k) =C_1 \rho\sb f \, \big[ \varepsilon
(k)\big / \rho\sb{eff}(k)\big]^{2/3}k^{-5/3}$.
Introduce the dimensionless ratio
\begin{eqnarray}
  \label{eq:ratio}
  R(k)\equiv \frac{E(k)\big / E(1/L) }{E_0(k)\big / E_0(1/L)}=
  \left[\frac{\ve(k)\rho_{_L}}
  {\ve_{_L}\rho\sb{eff}(k)}\right]^{2/3}\ .
\end{eqnarray}
Here $E_0(k)=C_1 \varepsilon^{2/3} \rho\sb f^{1/3}k^{-5/3}$ is the
density of turbulent kinetic energy in the particle-free case.  This
ratio is larger (smaller) than unity in the case of enhancement
(suppression) of the turbulent energy by particles. To understand this
behavior consider three distinct regions of scales defined by the
crossover scale $k_*$ ( see Table \ref{tab:1}):
$\B a$. \it{Region of small scales}, $L^{-1}<k< k_*$, where $\ve(k)$
is decreasing function of $k$. Function $\rho\sb{eff}(k)$ in the
denominator of \REF{ratio} is almost constant $\rho\sb f(1+\phi)$.

$\B b$. \it{Region of transient scales}, $k\sim k_*$, where $\ve(k)$
still decreases, and so does $\rho\sb{eff}(k)$ gradually reducing to
$\rho\sb f$.

$\B c$.  \it{Region of large scales}, $k > k_*>\eta^{-1}$, where
$\ve(k)$ approaches plato, while $\rho\sb{eff}(k)$ is again constant, 
$\rho\sb f$.  It is clear that behavior of $R(k)$ will depend on
$k_*L$.  For $k_*L\gg 1$ (small enough $\tau\sb p$) the ratio $R(k)$
has enough room in the region $\B a$ to strongly decay [as
$(kL)^{-2\Delta/3}$] due to decay of $\ve(k)$. In the region $\B b$ it
may increase [due to decrease of $\rho\sb{eff}(k)$] but not more than
by factor of $(1+\phi)^{2/3}$. Therefore for small enough $\tau\sb p$
the ratio $R(k)<1$ everywhere, see, \eg, in Fig.~\ref{f:3}a plots for
$\tau_p=0.05$,( $k_* L\approx 350$) and $\tau_p=0.1$ ($k_* L= 117$).

There is an essential enhancement of the kinetic energy for larger
$\tau\sb p$ (smaller $k_* L$), when the small scale region $\B a$ is
not pronounced (plots for $\tau\sb p=0.5$ ($k_*L=9.5$) and $\tau\sb
p=0.75$ ($k_*L=5.2$) in Fig.~\ref{f:3}a). In this case the growth of
$R(k)$ in the transient region $\B b$ is stronger than the decay in
the region $\B a$.

To find values of parameters for which  the enhancement is
possible (at least for $k\to\infty$) consider the maximum value of
$R_\infty=R(\infty)$.  In the second iteration
\begin{eqnarray}
  \label{eq:R2inf}
  R_{2,\infty}=\left(\frac{\delta}{1+\delta}\right)^{2
  \Delta/(2-\Delta)} \left(1+\frac{\delta
  \phi}{1+\delta+\phi}\right)^{2/3}\ .
\end{eqnarray}
Fig.~\ref{f:3}b displays plots $R_{2,\infty}$ \vs $\tau\sb p$ for
different $\phi$. The enhancement is possible if
$\delta=\tau\sb p\gamma_{_L}/(1+\phi)$ exceeds a critical value
$\delta\sb {cr}$ which is independent of $\phi$. Parameter $\phi$
governs the value of the enhancement. The maximal possible enhancement
(for large $kL$ and $\delta$) is $(1+\phi)^{2/3}$.

In is pleasure to acknowledge discussions with T. Elperin,
N. Kleeorin, G. Ooms, I. Procaccia, and I. Rogachevskii. This work
was supported by the Israel Science Foundation.


\begin{thebibliography}{99}

\bibitem{98CST} C.T. Crowe, M.Sommerfeld and Y.Tsuji, \it{Multiphase
flows with particles and droplets}, CRC Press, New York (1998).

\bibitem{83EA}  S.E. Eglobashi and T.W. Abou Arab,
Phys. Fluids  \textbf{26}, 931 (1983); 
O.A. Druzhinin and S.E. Eglobashi,
 Phys. Fluids, \textbf{11}, 602 (1999).

\bibitem{98BSS} M. Boivin, O. Simonin and K.D. Squires,
J. Fluid Mech. \textbf{ 375}, 235 (1998) 

\bibitem{00BSS} M. Boivin,
O. Simonin and K.D. Squires, Physics of Fluids, \textbf{8}, 2080
(2000).

\bibitem{94Egl} S.E. Eglobashi,
Appl. Sci. Res. \textbf{52}, 309 (1994).


\bibitem{00HKR} M. Hussainov, A. Kartushinsky, U. Rudi, I. Sheheglov,
G. Kohnen and M. Sommerfeld, 
Int. J. Heat Fluid Flow \textbf{21}, 365 (2000).

\bibitem {02LOP} V.S. L'vov, G. Ooms and A. Pomyalov, \it{Effect of
  partice inertia on the turbulence in suspension}, in preparation.



\end{thebibliography}
\end{document}